\documentclass[aps,prl,twocolumn,superscriptaddress,byrevtex,showpacs,preprintnumbers,amsmath,amssymb]{revtex4}

\usepackage{graphicx}
\usepackage{color}

\begin{document}

\title{Quantum reflection of helium atom beams from a microstructured grating}

\author{Bum Suk Zhao}
\affiliation{Fritz-Haber-Institut der Max-Planck-Gesellschaft,
 Faradayweg 4-6, 14195 Berlin, Germany}
\author{Stephan A. Schulz}
\affiliation{Universit\"at Ulm, Institut f{\"u}r
Quanteninformationsverarbeitung, Albert-Einstein-Allee 11, 89069
Ulm, Germany}
\author{Samuel A. Meek}
\author{Gerard Meijer}
\author{Wieland Sch\"ollkopf}
 \email{wschoell@fhi-berlin.mpg.de}
\affiliation{Fritz-Haber-Institut der Max-Planck-Gesellschaft,
 Faradayweg 4-6, 14195 Berlin, Germany}
\date{\today}

\begin{abstract} We observe high-resolution diffraction
patterns of a thermal-energy helium-atom beam reflected from a
microstructured surface grating at grazing incidence. The grating
consists of 10-$\mu$m-wide Cr strips patterned on a quartz substrate
and has a periodicity of 20 $\mu$m. Fully-resolved diffraction peaks
up to the $7^{\rm th}$ order are observed at grazing angles up to 20
mrad. With changes in de Broglie wavelength or grazing angle the
relative diffraction intensities show significant variations which
shed light on the nature of the atom-surface interaction potential.
The observations are explained in terms of quantum reflection at the
long-range attractive Casimir-van der Waals potential.
\end{abstract}


\maketitle

Optical elements, such as mirrors and coherent beam splitters for
matter waves, are prerequisite for atom and molecule interferometry.
Both gratings formed by laser light and material gratings have been
employed in Ramsey-Bord\'{e} and in Mach-Zehnder matter-wave
interferometers, respectively \cite{AI}. As the de Broglie
wavelengths of atoms and molecules at thermal energies are typically
$\leq 0.1$~nm, free-standing material transmission gratings of
sub-micron periodicity had to be used in interferometers for beams
of Na atoms \cite{Kei91b}, dimers \cite{Cha95}, and C$_{60}$
fullerenes \cite{Arn02,Arn07}. In addition, diffraction by a
100-nm-period transmission grating was applied to quantitatively
determine long-range atom-surface van der Waals potentials
\cite{ich:prl99,Bru02} and to investigate small He clusters
\cite{ich:science94}. Those gratings are, however, difficult to
make, expensive, and fragile. Shimizu and coworkers demonstrated
diffraction of ultracold atoms, released from a magneto-optical
trap, by a 2-mm-period surface grating with reflective strips
consisting of parallel 100-nm-wide ridges \cite{Shimizu02a}. Most
recently, partially resolved diffraction peaks of thermal beams of
metastable rare-gas atoms reflecting from a 2-$\mu$m-period surface
grating were reported \cite{Grucker07}.

Here, we present diffraction patterns of He atom beams that are
coherently reflected from a home-made 20-$\mu$m-period surface
grating under grazing incidence. For incident grazing angles in the
milliradian range the resulting diffraction angles are of the same
order of magnitude as the ones observed with a 100-nm-period
transmission grating at normal incidence \cite{ich:pra98}. The
projection of the grating period along the incident beam direction
yields an effective grating period in the sub-$\mu$m range. Yet, a
20-$\mu$m-period surface grating can readily be made out of a
variety of materials using standard lithographic techniques. Unlike
He atom beam scattering from smooth crystalline surfaces
\cite{Hulpke92}, ultra-high vacuum and {\it in-situ} surface
preparation are not needed, but coherent reflection is achieved with
a microscopically rough surface.

We present evidence for the underlying coherent reflection mechanism
being {\it quantum reflection} at the attractive long-range branch
of the atom-surface interaction \cite{HFriedrich02}. Quantum
reflection from a solid surface was observed recently with ultracold
metastable Ne \cite{Shimizu01} and He atoms \cite{Oberst05a}, with a
Bose-Einstein condensate \cite{Pasquini06}, and with a $^3$He atom
beam \cite{DeKieviet03}. It was described theoretically in terms of
the long-range Casimir-van der Waals atom-surface potential
\cite{HFriedrich02}. Furthermore, Shimizu and coworkers reported
diffraction of ultracold metastable He atoms quantum-reflected from
ridged surfaces \cite{Shimizu02a,Shimizu02b}.

\begin{figure}[b]
\includegraphics[scale=0.32]{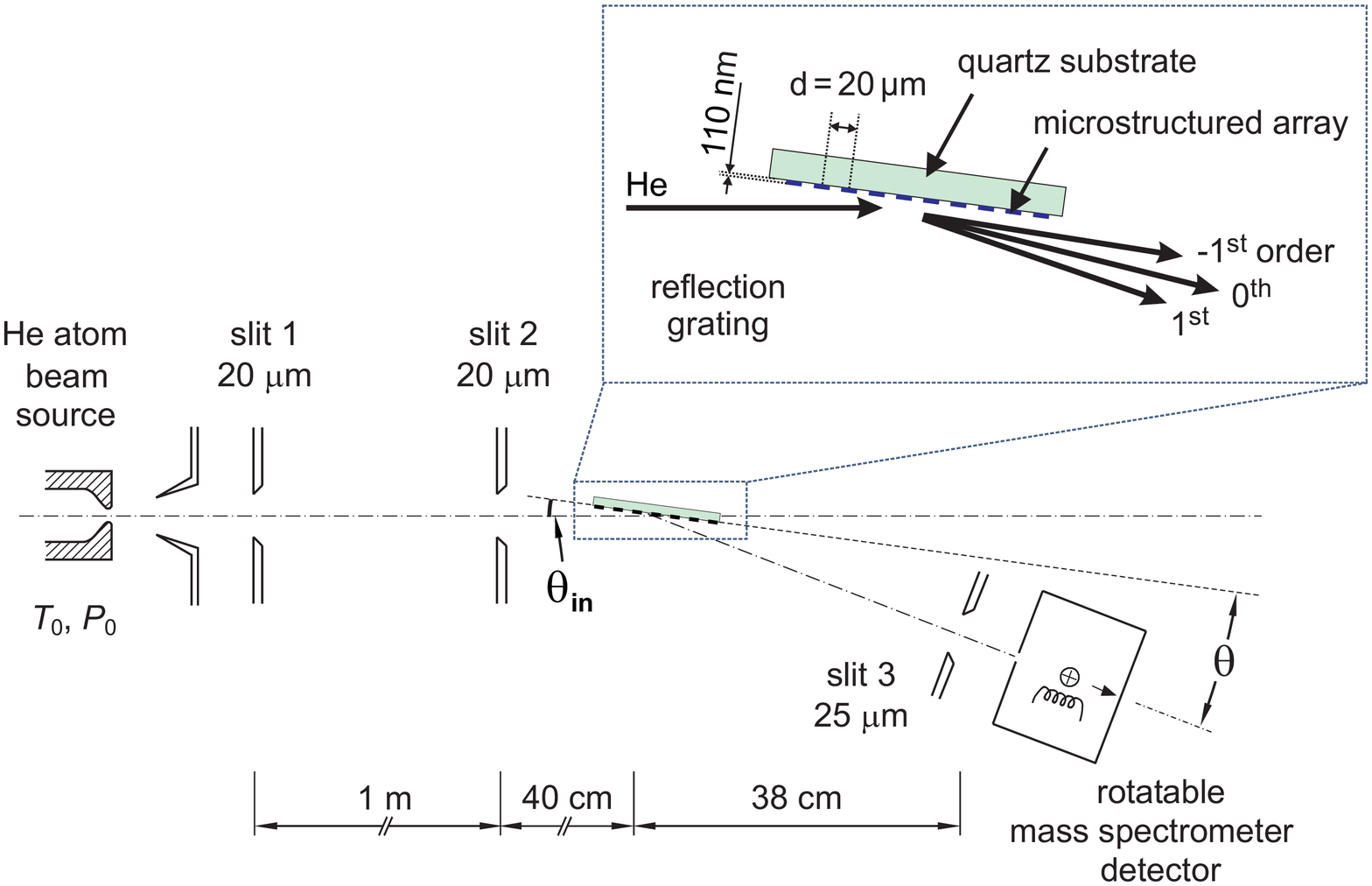}
\caption{(color online). Scheme of the experimental setup.
Diffraction patterns are recorded by scanning the detection angle
$\theta$ which is defined with respect to the reflection-grating
surface plane. The inset in the upper right shows an enlargement of
the grating indicating the directions of the 0$^{\rm th}$ and $\pm
1^{\rm st}$-order diffraction beams.} \label{fig:setup}
\end{figure}

In the apparatus \footnote{This apparatus was built at the
Max-Planck-Institut f\"ur Str\"omungsforschung in G\"ottingen,
Germany.} a helium atom beam is formed by free-jet expansion of pure
$^4$He gas from a source cell (stagnation temperature $T_0$ and
pressure $P_0$) through a 5-$\mu$m-diameter orifice into high
vacuum. As indicated in Fig.\ \ref{fig:setup} the beam is collimated
by two narrow slits, each 20~$\mu$m wide, located 15~cm and 115~cm
downstream from the source. A third 25-$\mu$m-wide detector-entrance
slit, located 38~cm downstream from the grating, limits the angular
width of the atomic beam to a full width at half maximum of
130~$\mu$rad. The detector is an electron-impact ionization mass
spectrometer that can be rotated precisely around the angle $\theta$
indicated in Fig.~\ref{fig:setup}. The microstructured reflection
grating is positioned at the intersection of the horizontal atom
beam axis and the vertical detector pivot axis such that the
incident atom beam impinges under grazing incidence (incident
grazing angle $\theta_{\rm in} \leq 20$ mrad), with the grating
lines oriented parallel to the pivot axis. The diffraction pattern
is measured by rotating the detector around $\theta$ and measuring
the mass spectrometer signal at each angular position.

The reflection grating consists of a 56-mm-long microstructured
array of 110-nm-thick, 10-$\mu$m-wide, and 5-mm-long parallel
chromium strips on a flat quartz substrate. It was made from a
commercial chromium mask blank by e-beam lithography. As shown in
the inset of Fig.\ \ref{fig:setup} the center-to-center distance of
the strips, and thereby the period $d$, is 20~$\mu$m. Given this
geometry the quartz surface between the strips is completely
shadowed by the strips for all the incidence angles used. We expect
a chromium oxide surface to have formed while the grating was
exposed to air before mounting it in the apparatus where the ambient
vacuum is about $8 \times 10^{-7}$~mbar. No {\it in-situ} surface
preparation was done.

\begin{figure}[pt]
\includegraphics[scale=0.41]{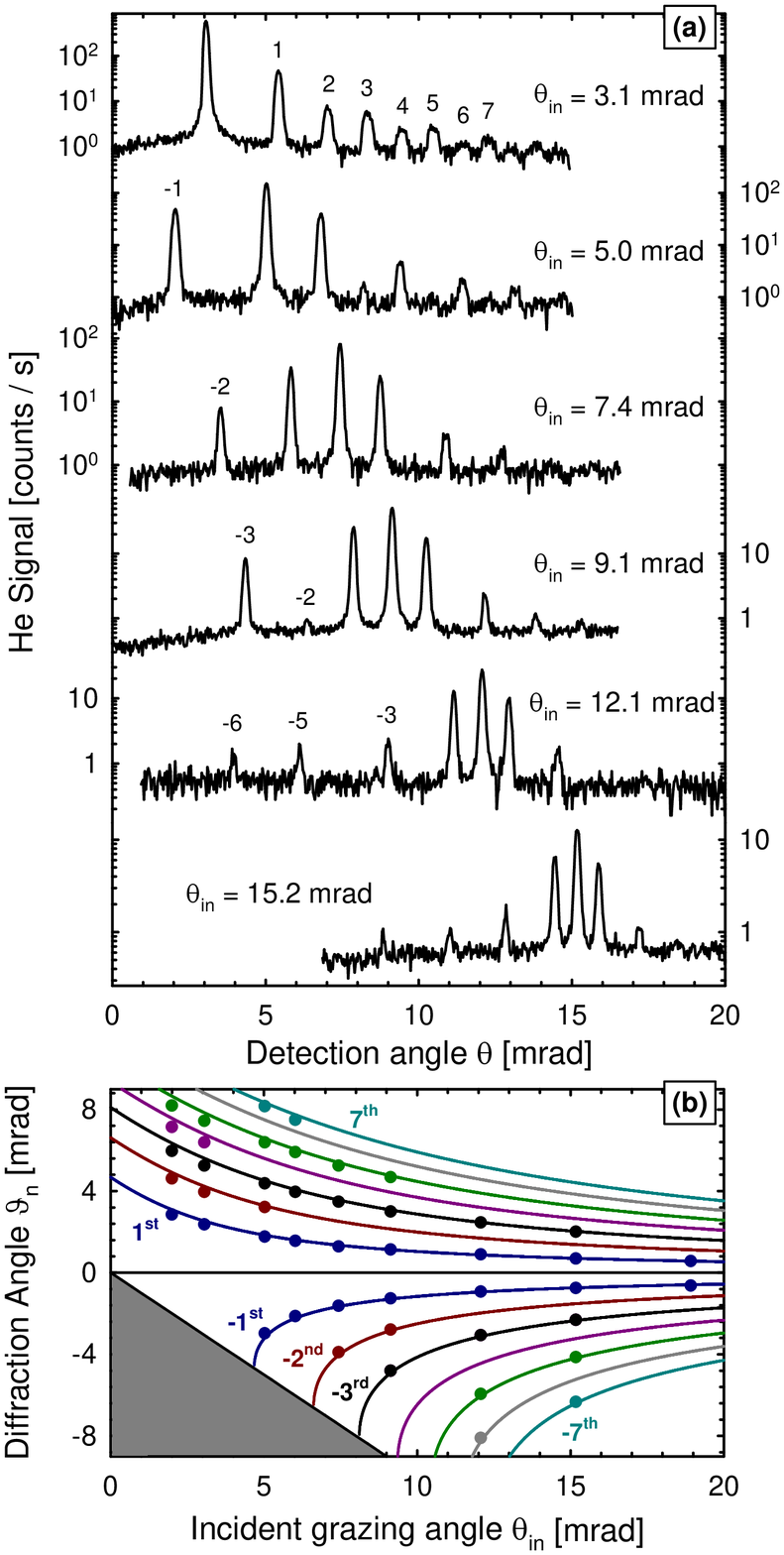}
\caption{(color online). (a) Semi-logarithmic plot of diffraction
patterns observed for a He atom beam at $T_0=20$ K at various
incident grazing angles as indicated. Numbers indicate the
diffraction-order assigned to the peaks. (b) Comparison between the
observed diffraction angles and those calculated by the grating
equation. The grey-shaded region indicates the regime beneath the
surface given by $\vartheta_n < - \theta_{\rm in}$.}
\label{fig:anglescan}
\end{figure}

Fig.~\ref{fig:anglescan}(a) shows a series of diffraction patterns
measured at constant source conditions of $T_0 = 20$~K and $P_0 =
6$~bar corresponding to a de Broglie wavelength of $\lambda =
2.2$~\AA. The incident grazing angle $\theta_{\rm in}$ was varied
between 3 and 15~mrad. The most intense peak in each diffraction
pattern is attributed to the specular reflection (0$^{\rm th}$
diffraction-order peak), for which the detection angle is equal to
the incident grazing angle. The intensity of the specular peak
decreases continuously from about 600 counts/s at $\theta_{\rm in}=
3.1$ mrad to only 13 counts/s at $\theta_{\rm in}= 15.2$ mrad. At
$\theta_{\rm in}= 3.1$ mrad at least seven positive-order
diffraction peaks can be seen at angles larger than the specular
angle (diffraction 'away from' the surface), while no negative
diffraction-order peak is present. With increasing incident grazing
angle negative-order diffraction peaks appear successively.

The diffraction angles $\vartheta_n$ are defined as the angular
separation between the $n^{\rm th}$ and $0^{\rm th}$
diffraction-order peak, $\vartheta_n = \theta_{n} - \theta_{0}$, and
are analyzed in Fig.~\ref{fig:anglescan}(b). It is straightforward
to calculate the diffraction angles using the grating equation
$\cos(\theta_{\rm in})-\cos(\theta_n)= n\frac{\lambda}{d}$
\cite{BornWolf59}. Here $\theta_n$ is the angle (with respect to the
grating surface plane) of the $n^{\rm th}$ diffraction-order peak,
and $\lambda$ is the de Broglie wavelength. The calculated
diffraction angles (lines in Fig.~\ref{fig:anglescan}(b)) agree with
the observed ones within the experimental error, thereby
unambiguously confirming the interpretation of the peaks as
grating-diffraction peaks. Note that with decreasing incidence angle
the negative-order diffraction peaks disappear successively when the
cut-off condition $\vartheta_n < -\theta_{\rm in}$ is met, i.e.,
when the peak is diffracted 'into the surface'. This regime is
indicated by the grey-shaded region in Fig.~\ref{fig:anglescan}(b).

The relative diffraction peak intensities change significantly with
incident grazing angle. For instance, for $\theta_{\rm in}= 3.1$
mrad even and odd order peaks have similar heights falling off
almost monotonously with increasing diffraction order. With
increasing incident grazing angle, however, the positive even-order
diffraction peaks tend to disappear. Moreover, a distinct
peak-height variation can be seen for the $-2^{\rm nd}$-order peak
which decreases sharply when $\theta_{\rm in}$ is increased from
$7.4$ to $9.1$~mrad.

\begin{figure}[pt]
\includegraphics[scale=0.47]{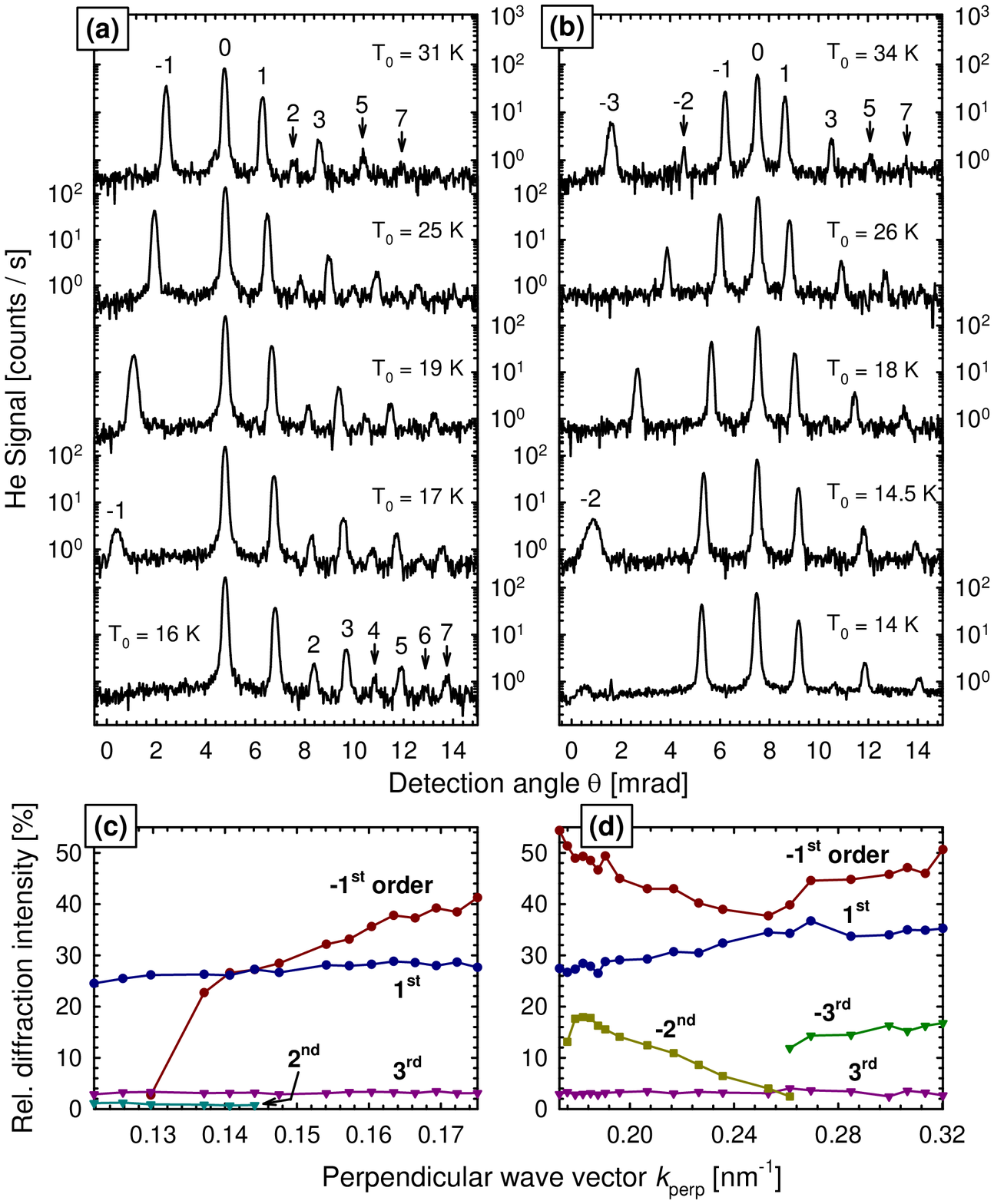}
\caption{(color online). Diffraction patterns observed for He atom
beams at various source temperatures for $\theta_{\rm in}= 4.9$ mrad
(a) and $\theta_{\rm in}= 7.2$ mrad (b). Numbers above the arrows
indicate the diffraction-order assigned to the peaks. Relative
diffraction intensities, that are normalized to the $0^{\rm
th}$-order intensity, are plotted for both series in (c) and (d),
respectively, as a function of $k_{\rm perp}$, the normal component
of the incident wave vector. The relative error of these data points
is around 10\%.} \label{fig:tempscan}
\end{figure}

To investigate the origin of these peak-height variations we studied
their dependence on the de Broglie wavelength by measuring two
series of diffraction patterns at constant incident grazing angles
of $\theta_{\rm in}= 4.9$ mrad and $\theta_{\rm in}= 7.2$ mrad (Fig\
\ref{fig:tempscan}). In each series we varied the source temperature
$T_0$ and, hence, the de Broglie wavelength which is approximately
proportional to $1/\sqrt{T_0}$ \cite{Miller88}. The broadening of
peaks at small $\theta$ is readily understood from
Fig.~\ref{fig:anglescan}(b) where it can be seen that a small width
of the incident $\theta_{\rm in}$ distribution leads to an increased
width in $\theta$ due to the steep slope of the lines close to the
cut-off.

To quantify the observed diffraction peak-height variations we
determine the relative diffraction intensities from the area under
each diffraction peak divided by the area of the corresponding
$0^{\rm th}$-order peak. In Fig.~\ref{fig:tempscan}(c) and (d) the
relative diffraction intensities are plotted as a function of
$k_{\rm perp}\approx \sin(\theta_{\rm in})\sqrt{5 k_B m T_0}/\hbar$,
the normal component of the incident wave vector (with $k_B$ the
Boltzmann constant, $m$ the particle mass, and $\hbar$ Planck's
constant over $2\pi$). The plots reveal (i) pronounced variations of
the diffraction intensities with $k_{\rm perp}$ for some diffraction
orders, (ii) even-order diffraction intensities up to 18\%, and
(iii) significant asymmetries between corresponding positive and
negative diffraction-order intensities. All three observations
cannot be explained by the Fraunhofer-Kirchhoff diffraction theory
which predicts the relative diffraction intensities for an amplitude
grating to be independent of wavelength and to be determined by the
ratio of grating period to strip width only \cite{BornWolf59}. As
this ratio is 2 for our grating, the even diffraction-order
intensities are expected to vanish altogether. This behavior is
indeed observed, but only in the limit of relatively large
$\theta_{\rm in}$ and $\theta$ as can be seen in
Fig.~\ref{fig:anglescan}.

We therefore attribute our observations to the long-range
atom-surface interactions which have been observed previously in
experiments with nanoscale transmission gratings
\cite{ich:prl99,Bru02,Cronin05}. For the reflection grating the
relative diffraction intensities are expected to be an even more
sensitive probe of the atom-surface interaction than for a
transmission grating because every part of the atomic wavefunction
probes the interaction potential. In addition, for sufficiently
small angles above the surface the interaction path length is
increased. A detailed theoretical model correctly accounting for the
phase shift induced by the surface potential would be needed to
determine the potential strength parameter from the data.

\begin{figure}[pt]
\includegraphics[scale=0.52]{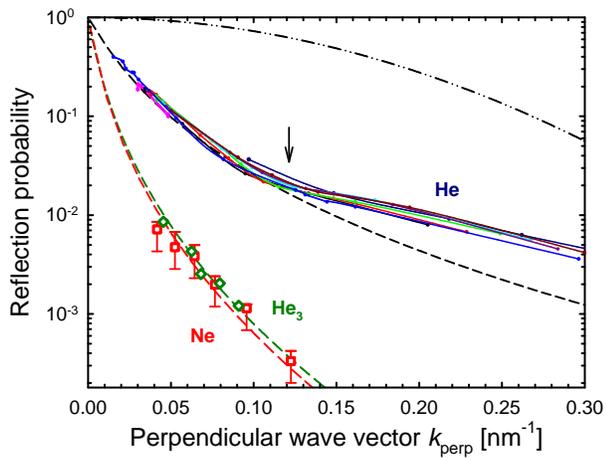}
\caption{(color online). Observed coherent reflection probabilities
for beams of He atoms (dots connected by solid lines, different
lines correspond to different $T_0$ (6.7 to 20~K)), He trimers (open
diamonds), and Ne atoms (open squares). The dash-dotted line
presents a prediction for {\it classical reflection} from a
hard-wall surface tentatively assuming a roughness $\sigma = 4$ nm.
The dashed lines present predictions for {\it quantum reflection} at
the long-range attractive branch of the surface potential. The kink
is marked by an arrow.} \label{fig:reflectivity}
\end{figure}

In Fig.~\ref{fig:anglescan} it can be seen that the peak heights
decrease with increasing incident grazing angle. The total coherent
reflection probability of the chromium strips is determined from the
sum of all peaks normalized to the incident beam signal and
multiplied by two to compensate for the 50\% chromium coverage of
the microstructure area. The reflection probability is as large as
40\% at $k_{\rm perp} = 0.015$~nm$^{-1}$ and decreases continuously
to less than 1\% at $k_{\rm perp} = 0.3$~nm$^{-1}$ (Fig.\
\ref{fig:reflectivity}). A kink at $k_{\rm perp} \simeq
0.12$~nm$^{-1}$ separates a steep decay at smaller $k_{\rm perp}$
from a slow simple exponential decay at larger $k_{\rm perp}$. The
dependence on $k_{\rm perp}$ differs significantly from what is
expected for specular reflection of a wave from a randomly-rough
hard-wall surface at grazing incidence, which is given by $\exp{(-(2
\sigma k_{\rm perp})^2)}$, where $\sigma$ denotes the
root-mean-square roughness of the surface \cite{Beckmann63}. While
this term does predict that even a rough surface reflects coherently
in the limit $\sigma k_{\rm perp} \rightarrow 0$, its dependence on
$k_{\rm perp}$ exhibits the wrong curvature as displayed by the
dash-dotted line in Fig.\ \ref{fig:reflectivity}.

A better agreement is obtained when we assume quantum reflection at
the long-range attractive branch of the atom-surface potential to be
the mechanism for coherent reflection. We calculate the quantum
reflection probability by numerically solving the 1-dimensional
Schr{\"o}dinger equation for an attractive Casimir-van der Waals
surface potential of the approximate form $V(z)=-C_4/\left[ (l+z)z^3
\right]$. Here, $z$ denotes the distance from the surface, and the
coefficient $C_4=C_3 l$ is the product of the van der Waals
coefficient $C_3$ and a characteristic length $l$ ($l = 9.3$~nm for
He) indicating the transition from the van der Waals ($z \ll l$) to
the Casimir regime ($z \gg l$) \cite{HFriedrich02}. For small
$k_{\rm perp}$ (left to the kink in Fig.~\ref{fig:reflectivity})
good agreement with the data is found for $C_3=2.5 \times 10^{-50}$
Jm$^3$. This value is slightly smaller than what is expected for He
interacting with a transition metal surface ($3.2 - 4.3 \times
10^{-50}$ Jm$^3$) but larger than what is expected for an insulating
surface \cite{Vid91}. We attribute this behavior to an insulating
chromium oxide surface having formed while the microstructure was
exposed to air.

The kink and the slow decay at larger $k_{\rm perp}$ are not
reproduced by this simple model. A similar kink was observed for He
reflecting from a liquid He surface at $k_{\rm perp} \simeq
0.35$~nm$^{-1}$ \cite{Nayak83} and was explained by a
quantum-reflection model based on a more realistic attractive
potential shape \cite{Edwards78}. As with increasing $k_{\rm perp}$
quantum reflection occurs at smaller $z$ \cite{HFriedrich02}, the
potential well region where the attractive branch deviates from a
simple $-C_3/z^3$ power law starts to influence the
quantum-reflection probability. Our calculations indicate that this
effect can be neglected only as long as $k_{\rm perp} <
0.1$~nm$^{-1}$ where we find agreement with the calculation.

Further support for quantum reflection is given by the observation
of reflected beams of He trimers and Ne atoms. For He trimers ($T_0
= 8.7$~K) the total coherent reflection probability increases by an
order of magnitude to almost 10$^{-2}$ when $k_{\rm perp}$ is
lowered from 0.09 to $0.045$~nm$^{-1}$
(Fig.~\ref{fig:reflectivity}). The data is modeled very well with a
$C_3$ coefficient 3 times that used for the He monomer calculation,
as one would expect for a van der Waals-bound cluster of 3 He atoms.
The binding energy of the He trimer is only 11 $\mu$eV \cite{Lew97}.
Hence, it is more than three orders of magnitude smaller than the
well depth of the estimated He$_3$-surface potential. As a result,
classical reflection at the potential's repulsive inner branch
should inevitably lead to dissociation.

For a Ne atom beam ($T_0 = 40$~K) the observed reflectivity data is
well matched by the calculated quantum-reflection probability for
$C_3=4.0 \times 10^{-50}$ Jm$^3$ (dashed line in
Fig.~\ref{fig:reflectivity}, characteristic length $l = 11.84$~nm).
As in the case of He this $C_3$ coefficient is again somewhat
smaller than expected for a transition metal ($7 - 9 \times
10^{-50}$~Jm$^3$ \cite{Vid91}).

In summary, we observed fully resolved diffraction peaks up to the
$7^{\rm th}$ order for a He atom beam reflected from a
20-$\mu$m-period microstructured surface grating under grazing
incidence. The observed relative diffraction intensities vary
significantly with incident grazing angle and de Broglie wavelength.
Furthermore, we observed coherent reflection probabilities from the
surface grating for beams of Ne and of He$_3$ as a function of the
normal component of the incident wave vector. The measurements are
in excellent agreement with the predictions from a simple
1-dimensional quantum-reflection model.

\begin{acknowledgments} B.S.Z.\ acknowledges support by the Alexander
von Humboldt Foundation and by the Korea Research Foundation Grant
funded by the Korean Government (KRF-2005-214-C00188). We thank R.
Br\"uhl for assistance with the data-acquisition software and
H.~Conrad and J.R.~Manson for fruitful discussions.
\end{acknowledgments}

\end{document}